\documentclass[10pt,preprint]{emulateapj}

\begin{document}

\title{A large number of $z>6$ galaxies around a QSO at $z=6.43$:
Evidence for a protocluster?} \thanks{Based  on data collected at Subaru Telescope, which is operated by the National Astronomical Observatory of Japan.}
\author{Yousuke Utsumi}
\affil{The Graduate University for Advanced Studies, 2-21-1 Osawa Mitaka Tokyo, Japan}
\affil{National Astronomical Observatory of Japan, 2-21-1 Osawa Mitaka Tokyo, Japan}
\email{yousuke.utsumi@nao.ac.jp}
\author{Tomotsugu Goto}
\affil{Institute for Astronomy, University of Hawaii}
\affil{Subaru Telescope 650 North A'ohoku Place Hilo, HI 96720, USA}
\author{Nobunari Kashikawa}
\author{Satoshi Miyazaki}
\author{Yutaka Komiyama}
\affil{The Graduate University for Advanced Studies, 2-21-1 Osawa Mitaka Tokyo, Japan}
\affil{National Astronomical Observatory of Japan, 2-21-1 Osawa Mitaka Tokyo, Japan}
\author{Hisanori Furusawa}
\affil{National Astronomical Observatory of Japan, 2-21-1 Osawa Mitaka Tokyo, Japan}
\author{Roderik Overzier}
\affil{Max-Planck Institut f\"ur Astrophysik, D-85741 Garching, Germany}

\begin{abstract}
  QSOs have been thought to be important for tracing highly biased
  regions in the early universe, from which the present-day massive
  galaxies and galaxy clusters formed.  While overdensities of
  star-forming galaxies have been found around QSOs at $2<z<5$, the
  case for excess galaxy clustering around QSOs at $z>6$ is less
  clear.  Previous studies with HST have reported the detection of
  small excesses of faint dropout galaxies in some QSO fields, but
  these surveys probed a relatively small region surrounding the
  QSOs. To overcome this problem, we have observed the most distant
  QSO at $z=6.4$ using the large field of view of the Suprime-Cam
  ($34'\times 27'$).  Newly-installed red-sensitive fully depleted CCDs allowed us to
  select Lyman break galaxies (LBG) at $z\sim6.4$ more efficiently. We
  found seven LBGs in the QSO field, whereas only one exists in a
  comparison field.  The significance of this apparent excess is
  difficult to quantify without spectroscopic confirmation and
  additional control fields. The Poisson probability to find seven
  objects when one expects four is $\sim$10\%, while the probability
  to find seven objects in one field and only one in the other is less
  than 0.4\%, suggesting that the QSO field is significantly overdense
  relative to the control field. These conclusions are supported by a
  comparison with a cosmological SPH simulation which includes the
  higher order clustering of galaxies.  We find some evidence that the
  LBGs are distributed in a ring-like shape centered on the QSO with a
  radius of $\sim$3 Mpc. There are no candidate LBGs within 2 Mpc from
  the QSO, i.e., galaxies are clustered around the QSO but appear to
  avoid the very center.  These results suggest that the QSO is
  embedded in an overdense region when defined on a sufficiently large
  scale (i.e. larger than an HST/ACS pointing). This suggests that the
  QSO was indeed born in a massive halo. The
  central deficit of galaxies may indicate that (1) the strong UV
  radiation from the QSO suppressed galaxy formation in its vicinity,
  or (2) that star-formation closest to the QSO occurs mostly in an
  obscured mode that is missed by our UV selection.
\end{abstract}

\section{Introduction}

At $2<z<5$, strong overdensities of star-forming galaxies have been
found around QSOs and radio galaxies, and thus, QSOs have been thought
to trace highly biased regions in which the present-day massive galaxy
clusters formed
\citep[e.g.][]{2003ApJ...596...67D,2004Natur.427...47M,2007ApJ...663..765K}.
It is expected that this extends to the most luminous $z\sim6$ QSOs as
well, as this rare population hosts supermassive black holes of
several billion solar masses that are presumed to reside in the most
massive galaxies and dark matter halos present at this redshift.

However observational results to date have been puzzling; Five
$z\sim6$ QSO fields observed by the HST/ACS show no major enhancements
in the galaxy density \citep{2009ApJ...695..809K}.  Even though a few
QSO fields have showed an apparent overdensity
\citep{2005ApJ...622L...1S,2006ApJ...640..574Z}, none of them were
among the richest structures discovered to date at $z\sim6$, as
evidenced by much larger overdensities found in random fields
\citep[e.g., ][see Overzier et al. (2009) for a
discussion]{2004ApJ...611..685O,2007ApJ...663..765K,2008ApJ...677...12O}.
  
Why do QSOs suddenly appear to stop being at the center of the
overdensity at $z\sim6$?  One hypothesis is that despite the higher
dark matter densities near the QSOs, the strong ionizing UV radiation
from the central QSOs may prohibit the condensation of gas thereby
suppressing galaxy formation around the QSOs
\citep{1999ApJ...523...54B}. An alternative and perhaps more likely
explanation is that the lack of overdensities identified is related to
the fact that it is currently technically challenging to perform a
survey deep enough to detect faint $z\sim6$ galaxies and cover an area
large enough to probe the large-scale structure surrounding the QSOs.

In this work, we aim to study the large scale structure around the
currently most distant QSO at $z=6.43$, taking advantage of the large
aperture of the 8.2m telescope ``Subaru'' located on Maunakea
Observatory and the wide field prime focus camera ``Suprime-Cam''
(34'$\times$27') \citep{2002PASJ...54..833M}.  In addition, we have
recently installed new red-sensitive fully depleted CCDs on the Suprime-Cam
\citep{2008SPIE.7021E..52K}, allowing us to reach necessary depths in
much shorter exposure time.  Improved sensitivity is a factor of about
1.4 and 2 better in the $z'$-band and at $\sim 1\mu \rm m$,
respectively.
 
Throughout the paper, we use $H_{0} = 70h_{70}$ km s$^{-1}$ Mpc$^{-1}$, $\Omega_{m} = 0.3$, 
$\Omega_{\Lambda}= 0.7$. Magnitudes are given in the AB system. 

\section{Data \& analysis}

\subsection{Observation}
 
We observed the field surrounding the most distant QSO,
CFHQSJ2329-0301 at $z=6.43$, $M_{1450}=-25.23$
\citep{2007AJ....134.2435W} using Suprime-Cam with the filters
$i',z',z_R$ during an engineering run in Aug 2008 and UH time in Jun
2009.  We used the special $z_{R}$ band filter constructed by
\citet{2005PASJ...57..447S}. This filter covers the redder side of the
$z'$ band and has a central wavelength of $9900{\rm \AA}$
(Fig.\ref{fig:spectrum}).  The seeing was stable throughout the runs,
with a FWHM $\approx 0.5$ arcsec in $i'$, $0.4\sim 0.7$ arcsec in $z'$
and $0.5\sim 0.7$ arcsec in $z_{R}$. The exposure times were 360 and
500 s in $i'$ and $z_R$, respectively.  The exposure times in the $z'$
band varied; we obtained 3 exposures of 100 s, 2 of 300 s, 3 of 400 s,
6 of 500, and 2 of 700 s resulting in more than 10000 ADU of sky
counts.  The total exposure time in $i', z'$ and $z_{R}$ were 3,600 s
on Aug 28 2008, 6,900 s on 2, Aug 27 and 12,532 s on Aug 27 2008 and
Jun 18 2009 respectively, under good seeing conditions (less than 0.6
arcsec FWHM).
Our dither strategy consisted of five or more pointings on a circle of 1 arcmin radius.

\subsection{Data Reduction}

Our reduction procedure follows \citet{2002ApJ...580L..97M}, with
small adjustments made related to the new CCD.  A major difference
between the old and new Suprime-Cam is that the latter has 4 channel
read out circuits for faster readout.  As a result, over-scan
subtraction and background sky-subtraction should be performed on each
channel individually for each CCD.  An overscan value is evaluated by
taking the mean along the columns of the overscan region of each
channel, and then subtracted from the science region.  A flat field
image is created by taking the median of all science images after
masking objects.
The background sky is subtracted on each channel
after flat fielding. 
To estimate the background, we first reject data points having 2 times
larger or 4 times lower sigma value
than the sky value. The upper cut prevents astronomical
objects from contaminating the sky, while the
lower cut removes residuals from the flat-fielding.
Then, we estimate the modes in 64 pixel $\times$ 64 pixel boxes.
A background of each pixel is estimated by interpolating these mode
values at vertexes of a triangle centered on the pixel.

Before combining all the science frames, the images need to be
corrected for (1) geometric distortion, (2) displacement and rotation
of each CCD from the fiducial position, and (3) pointing offsets
between exposures.  We corrected for these effects by minimizing the
positional differences of common control stars identified on each CCD
in all exposures relative to the first exposure.  The resulting
positional error of the control stars from the fiducial point is
approximately 0.5 pixel (rms).  We solve the transformation among the
3 bands, aligning each frame with the first exposure of the $z_{R}$
band in order to perform multi-band photometry using an aperture as
small as possible to obtain a high S/N ratio.  Flux offsets are also
calculated during this procedure.  Since we want to determine the
transformation among the bands as accurately as possible, an image
warping procedure was performed only once, and a second order
bi-linear interpolation was used for rebinning. For stacking, we
adopted the clipped mean in order to eliminate cosmic rays.  The
implementation of these reduction steps were performed using the
software suite \emph{imcat}.

\subsection{Photometric Calibration}

Photometric calibration is performed by comparing stars in a reference
field 2 deg away to the North $(\alpha, \delta)_{2000}=(23:29,-01:01)$
with those in the SDSS/DR7 star catalog \citep{2008arXiv0812.0649A}
and using the following equations:
\begin{eqnarray}
	i'_{Subaru}&=&0.125(i'-r')_{SDSS}+0.003+i'_{SDSS}\\
	z'_{Subaru}&=&-1.091(i'-z')_{SDSS}+0.004+i'_{SDSS}\\
	z_{R,Subaru}&=&-1.414(i'-z')_{SDSS}+0.021+i'_{SDSS}
\end{eqnarray}
These equations are determined from \citet{gunn83} and convolved with
response curves that include both optics and the atmosphere. We
measured the efficiency of the new CCDs in
\citet{2008SPIE.7021E..52K}.  Since the calibration field was observed
soon after the completion of our science exposure, no
airmass/atmospheric corrections were needed to obtain the photometric
zero-point. After the correction, the rms of the magnitude differences
between our catalog and the SDSS catalog are 0.06, 0.07 and 0.07 in
$i', z'$ and $z_{R}$, respectively.

We have checked the colors of a sample of star-like objects selected
according to {\tt CLASS\_STAR>0.9}, by comparing the color of the
stars from \citet{gunn83} to the observed ones, confirming a good
internal consistency of colors in our catalog.  The template and
observed colors were consistent within an accuracy of 0.05 mag.  Since
the seeing in the final images are is 0.58, 0.54 and 0.50 arcsec
(FWHM), we use a small aperture to perform photometry to obtain a
better S/N ratio.  Following \citet{2005PASJ...57..447S}, who used an
aperture size twice that of PSF, we adopt a 1.2 arcsec aperture to
perform photometry.  The 3 sigma limiting magnitudes within this 1.2
arcsec aperture are $i' =26.95$, $z'=26.13$, and $z_{R} =25.46$ (AB)
mag.  Here, 1 $\sigma$ sky magnitudes are computed by randomly placing
a 1.2 arcsec aperture in the blank (sky) part of each image
\citep{2002AJ....123...66Y,2004ApJ...611..660O}.

\subsection{Astrometric Calibration}

We obtain an astrometric solution by comparing our catalog with the
USNO-B1.0 catalog (see \emph{scamp} \citep{2006ASPC..351..112B}).  The
resulting accuracy of the astrometric solution is 0.5 arcsec (rms).
Note that we did not use this astrometric solution to stack individual
raw images, but only to obtain the final absolute positions of the
objects.

\subsection{Object Detection}

The object detection is performed by SExtractor 2.3.2
\citep{1996A&AS..117..393B}.  We constructed a detection image, the
``all''-$z_{R}$ image, by combining all science frames, including
those with a slightly poorer seeing conditions.  The object detection
reliability using this ``all''-$z_{R}$ image is higher than when only
using the $z_{R}$ image for detection given that $\approx3\sigma$
sources in the $z_{R}$ image are detected at $\approx3.7\sigma$ in the
``all''-$z_{R}$ image.  Using the ``all''-$z_{R}$ image, we consider
an object detected if it has more than 5 connected pixels with each
exceeding the local sky rms by a factor of 2.  To reduce contamination
by false detections, we only use the ``detected cleanly'' and ``no
blending'' objects that have {\tt FLAGS=0} in SExtractor.
We measure the magnitudes in 1.2 arcsec apertures for each pass-band
to derive colors of the detected objects, and adopted {\tt MAG\_AUTO}
as our estimate for the total $z_{R}$ magnitude.  Objects in the
$i'$-band fainter than 2 sigma were replaced by the 2 sigma limiting
magnitude as an upper limit. In addition, all objects having
detections in the $z'$-band of less than 2 $\sigma$ were rejected in
order to keep spurious detection at a minimum.

Some lower S/N regions, such as the ``blooming regions'' (the halos
and horizontal spikes surrounding bright stars), as well as the outer
edge of the image, are masked manually.  The resulting effective field
of view is 0.219 deg$^{2}$.  We detect 48,632 objects after masking
(to a limiting magnitude of 25.46, or 3.7 sigma detections on the
``all''-$z_{R}$ image).  The number of detections is comparable to
that in our comparison field, the Subaru deep field (SDF), which
contains 45,405 objects in a single field of view of the Suprime-Cam
\citep{2005PASJ...57..447S}.

\subsection{SED Modeling}

\begin{figure}
   \centering
   \includegraphics[width=2.3in,angle=-90]{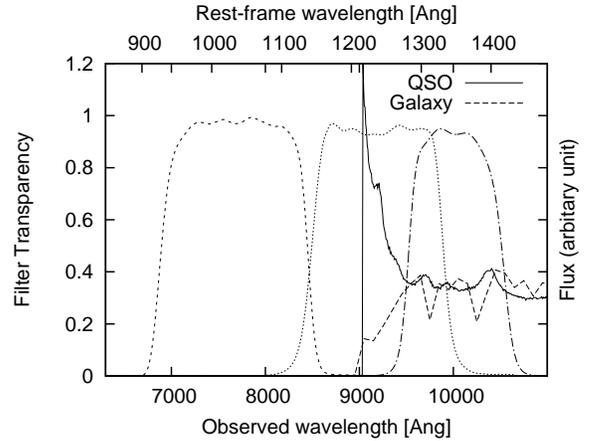} 
   \caption{An example of the spectra of a galaxy and a QSO redshifted to $z=6.43$. 
    The filter response curves in $i'$, $z'$ and $z_{R}$ are indicated from left to right. 
The spectrum of the galaxy is based on a model having an 
     exponentially-declining star formation history and an age of 0.5 Gyr \citep{2003MNRAS.344.1000B},
     while the spectrum of the QSO is based on the QSO composite spectrum taken from
     \citep{2001AJ....122..549V}.}
   \label{fig:spectrum}
\end{figure}

We have upgraded the CCDs of the Suprime-Cam to red-sensitive ones,
which have 1.4 and 2 times better sensitivity in $z'$ and $z_R$
compared to the previous detectors. Thus, we can reach the necessary
depth in much shorter exposure time.  This improved sensitivity allows
us to efficiently use the $z'-z_R$ color to select Lyman break
galaxies at $z>6.4$. At the QSO redshift of $z=6.43$, the $z'-z_R$
color selection is more efficient in selecting galaxies than the
$i'-z'$ color selection (the so-called $i$-dropout technique), which
has been used in previous works to select $5.5<z<6.5$ galaxies
(Fig.\ref{fig:spectrum}).

\begin{figure}
   \centering
   \includegraphics[width=3in]{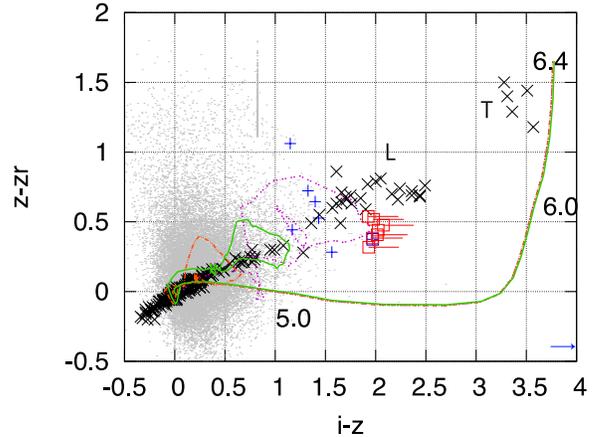} 
   \caption{$i'-z'$ vs $z'-z_{R}$ color-color diagram, showing the
     locations of detected objects, candidate $z\sim 6.4$ galaxies,
     stellar objects, and tracks indicating the expected colors of
     low- and high redshift galaxies.  Gray dots are all objects
     fainter than $z_{R}=24.0$, Red squares are $z\sim6.4$ galaxies
     selected in the QSO field.  Blue plusses are objects detected as
     $z\approx6$ galaxies in \citet{2005PASJ...57..447S} in the
     SDF. Gray box crosses are stars
     \citep{gunn83,2006AJ....131.2722C,2004AJ....127.3516G,2004AJ....127.3553K}.
     The purple, green, and orange lines represent model galaxies
     having star ``instantaneous burst'', ``exponential'' and
     ``constant''  star formation histories \citep{2003MNRAS.344.1000B}. Absorption by the
     neutral hydrogen at each redshift was added following the prescription of 
     \citet{2006ARA&A..44..415F}.  The blue arrow indicates CFHQS J2329-0301
     (see Table \ref{tab:qsolist}).}
   \label{fig:colorcolor}
\end{figure}

In order to predict the colors of galaxies as a function of their
redshift, we computed $z'-z_R$ colors of model galaxies (0.05, 0.1,
0.2 Gyr of age, using \citet{2003MNRAS.344.1000B}) and added the
absorption effects from the neutral hydrogen in the intergalactic
medium \citep{2006ARA&A..44..415F}.  According to our modeling, colors
of $i-z'>1.9$ and $z'-z_{R}>0.3$ can be used to identify galaxies at
$z>5.8$ (Fig. \ref{fig:colorcolor}).

\subsection{Object Selection}\label{selection}

Here we will select LBGs at $z>6.4$ using our $i',z',z_R$ catalog.
The model colors shown in Fig. \ref{fig:colorcolor} predicts that
galaxies at $z=6.4$ should have a large color difference of about
$i'-z'\simeq3.5$.  However, for an object having $z'\simeq25$ this
would require $i'\simeq28.5$ mag, much fainter than our detection
limits.  Therefore, we slightly loosen the color cut and require
$i'-z'>1.9$ and $z'-z_{R} > 0.3$. These relaxed criteria are still
appropriate for selecting LBGs at $z>6.4$, although we must be
somewhat cautious of potential galaxies at $z\simeq1.8$ having colors
of $i'-z'\simeq1.9$ that could make it into the sample as well.  In
order to assess the impact of this effect on our conclusions, in
section \ref{redgalaxies} below we will try to statistically estimate
the number of such $z\simeq 1.8$ interlopers expected.

In our control field (the SDF), only a single object was found that
satisfied our color criteria, and this object was classified as a
genuine $z\sim6$ galaxy by \citet{2005PASJ...57..447S}.  Because the
SDF data are deeper than ours, we adjusted the depth of that field by
replacing the magnitudes of all objects fainter than our limits with
the limiting magnitudes as computed for our field.
We also removed two objects brighter than $z_{R,Auto}=24.0$ since such
objects are too bright to be at $z=6.4$ (according to a $z\sim6$ LBG
study in the SDF, there is no $z>6$ galaxy brighter than $z_{R}=24.8$,
see \citet{2005PASJ...57..447S}).  These two objects may be
contaminating stars since they have SExtractor {\tt CLASS\_STAR}
values greater than 0.95, meaning that they are likely to be
unresolved.

As a result, to the magnitude limit of $z'<25.46$, we have detected 7
objects that are good candidates for being $z\sim6$ LBGs. These
objects are shown as red squares in Fig. \ref{fig:colorcolor}. Note
that the (lower limits on) their $i-z$ colors are bluer than expected
from the models, due to the relatively shallow limiting magnitude in
$i'$.

\section{Sources of Contamination}

\subsection{False Detections}

Because we are selecting very faint objects, our sample may be
contaminated by false detections due to background fluctuations.  To
evaluate the number of such contaminations, we created a negative of
the ``all''-$z_{R}$ image, and repeated our detection, photometry and
masking routines on each of the filter images. No detections were
obtained based on this negative image.  We conclude that our catalog
is not affected by spurious detection.

\subsection{Contamination by Faint Dwarf Stars}

As we can see in Fig. \ref{fig:colorcolor}, L/T dwarf stars also
satisfy our adopted color cut ($i'-z'>1.9$ \& $z'-z_{R}>0.3$).  In
this subsection, we estimate the expected number of L and T dwarfs, given
the size, depth and Galactic position of the field.

As a first test, when we applied our color criteria to the SDF only
one object was found and classified as a genuine $z\sim6$ object (see
above). The second reddest object in the field has $i-z\sim1.65$,
i.e., there are no stellar-like objects in the SDF with colors red
enough to make it into our selection.

However, we need to take into account the different galactic latitudes
of our field and the SDF.  Unfortunately, the late-type star counts as
a function of Galactic position are not accurately known at the faint
magnitudes we are probing. Late-type stars having similar colors to
high-$z$ galaxies can only be distinguished by using deep
spectroscopic observations.  The latest estimate was performed by
\citet{2008A&A...488..181C}, who studied the contamination from L, T
dwarfs by modeling the spatial distribution of late type stars in the
Galactic thin disc described using an exponential law
$n_{i}=n_{A,i}e^{-\frac{d}{d_{B}(l,b)}}, \frac{1}{d_{B}}\equiv
\frac{\cos b\cos l}{h_{R}}\pm\frac{\sin b}{h_{Z}}$, where $n_{A,i}$ is
the number density of $i$-type stars, $h_{R}$, $h_{Z}$ are the scale
length and the scale height of the exponential disc and $d$ is the
distance to us in galactic coordinates,
$d=(R_{\odot}^{2}+d^{2}\cos^{2}b -2R_{\odot} d \cos b \cos l)^{1/2}$.
The adopted parameters are $R=8.6 \rm kpc$, $h_{R}=2.25 \rm kpc$,
$h_{Z}=0.3\rm kpc$.  Local spatial densities of late type stars are
adopted from a compilation given in \citet{2008A&A...488..181C}, while
relations between the type and absolute magnitude are taken from
\citet{2004AJ....127.3553K}.  According to this model calculation, we
expect 1.5 times as many L/T dwarfs as in the SDF for $z_{R}<25.25$
mag, corresponding to the faintest bin of the magnitude of our
selected objects.

We have seen that there were no stars in our $z\sim 6.4$
color-selection in the SDF. Here we try to scale that null detection
in the SDF to our field.  If we assume that stars follow a Poisson
distribution, the expected number of L/T dwarfs in the SDF is 0.40 per
field at a 68\% confidence level.  Using the Galactic scaling computed
above, we expect that the probability of finding two or less stars in
our field is $87\%$.  Given that we have already rejected two stellar
like objects ($\S$2.5), the chance of finding one more star is less
than two\%.  Therefore, we conclude that there is little chance that
our $z\sim6.4$ LBG catalog is contaminated by stellar objects.  We
note, however, that object with id 1 could be a dwarf star since its
$i'-z'$ color is not large enough. We will include this object in the
following discussion.

\subsection{Contamination by red galaxies at $z\sim1.8$}
\label{redgalaxies}

As mentioned earlier, we found only one object when applying our
adopted color criteria for z$\sim$6.4 galaxies to the SDF catalog
(after adjusting the limiting magnitudes to ours), and this object was
identified as a $z\sim6$ object given that its $B, V, R$ magnitudes
were too faint for a low-$z$ interloper. Based on this, the expected
number of $z\sim1.8$ galaxies found according to our criteria should
be close to zero with an upper limit of 1.83 at a confidence level of
$84\%$ \citep{1986ApJ...303..336G}.  Although cosmic variance should
ideally be taken into account as well, we do not have any other
suitable comparison fields that are as deep in $z_{R}$ as the SDF.

However, we use the publicly available SXDS DR1 catalog
\citep{2008ApJS..176....1F} to obtain a rough estimate of the expected
contamination using only a $i-z>1.9$ color cut.  The SXDS covers 5
pointings with the Suprime-cam, and thus, the cosmic variance is less
of a concern.  We first applied again the observational limits as
given by our field by replacing all magnitudes that are fainter than
our limiting magnitudes ($i'_{2\sigma}=27.39, z'_{2\sigma}=26.57$) by
our 2$\sigma$ upper limits.  We used {\ttfamily FLAG=0} in order to
select only cleanly detected objects, and trimmed a region of lower
S/N (500 pixels from the edge of the field). Only 3 objects were found
that passed the selection criteria in all 5 pointings of the SXDF.
These objects are detected in the $B$ band (brighter than 28.4,
$3\sigma$ mag), indicating that they are most likely low redshift
interlopers at a redshift of $z\sim1.8$. If we scale this number to
the area of our single pointing field, only 0.6 of such interlopers
are galaxies are expected in our. This rough estimate is consistent
with the estimate derived above based on the SDF, suggesting that the
contamination from $z\sim1.8$ galaxies is negligible.

\section{Results}

\begin{figure}[tbp] 
   \centering
   \includegraphics[angle=-90,width=3.5in]{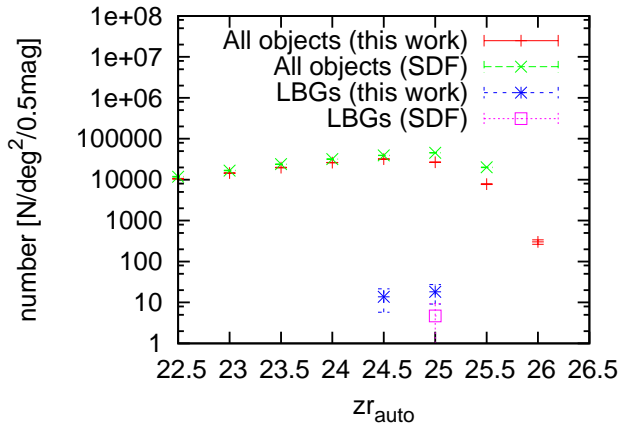}
   \caption{Number counts of all detected objects and the $z$-dropout galaxies.  We 
     did not apply completeness corrections.  The number of LBGs in our
     field is larger than that in the SDF although the opposite is true when comparing the total number of l
     detected objects.}
   \label{fig:numbercount}
\end{figure}

\subsection{Number Counts}

Fig. \ref{fig:numbercount} shows the number counts of all detected
objects, as well as those of the $z\sim6$ galaxies.  The number of all
objects in this field is slightly lower, by about $20\%$, than those
detected in the SDF field at $z_{R}<24$.  This could be due to a small
difference in the absolute photometric calibrations between the two
catalogs, or due to cosmic variance.  If it were due to a difference
in photometric calibration, a shift of 0.2 magnitude would be required
to explain the difference. Note that such an offset, if real, would
not affect the colors used in our color selection of the $z\sim 6$
galaxies, as the colors were checked for internal consistency using a
catalog of stars. We next consider the possibility of cosmic variance.
\citet{2008ApJS..176....1F} observed five Suprime-Cam pointings and
derived the number counts.  The variance in the number counts was
found to amount to a factor of 1.7 among the individual pointings in
the $z'$ band. Thus, we conclude that the differences in the number
counts found between our field and the SDF are consistent with being
due to cosmic variance.  Finally, we note that at magnitudes fainter
than $z_{R}=25$ the difference in number counts becomes larger again,
but this is simply because our data are slightly shallower than the
SDF.

\subsection{Comparing the LBG number density in the QSO field with that in the SDF}

We detect 7 objects as $z\sim6.4$ galaxy candidates using the color
criteria in $\S$ \ref{selection}, while 9 objects are rejected as
lower redshift objects at $z<5.5$ based on a $z'-z_{R}$ color cut (it
would have been difficult to eliminate these objects using only
$i'-z'$). For comparison, we apply the same color criteria to the SDF
catalog after applying our magnitude limits and find only one
candidate. Thus the number of $z\sim 6.4$ galaxies in the QSO field is
7 times larger than that the SDF field. We note that if we apply a
more relaxed cut based on a single color of $i-z>1.9$ (the standard
$i$-dropout technique) to the reference field (SDF), we still find 4
objects although 16 objects found in the QSO field,
implying that even with the standard $i$ dropout technique we
find an overdensity in the QSO field.

\begin{figure}[tbp]
   \centering
   \includegraphics[width=3.2in]{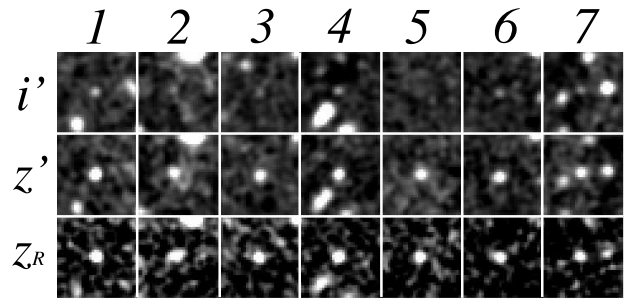} 
   \caption{Thumbnail images of the LBG candidates in the QSO
     field. The properties of the individual sources are listed in
     Table \ref{tab:objectlist}.}
   \label{fig:thumbs}
\end{figure}

The properties of the objects identified are summarized in Table
\ref{tab:objectlist}, and we show the filter thumbnail images in
Fig. \ref{fig:thumbs}. Although our data are shallower than the SDF,
we have detected more LBGs than in the SDF.  The result suggests that
the number density of $z\sim 6.4$ galaxies in this QSO field is larger
than that in the SDF field.  Such an overdensity would be consistent
with the predictions from cosmological simulations (e.g.,
\citet{2005Natur.435..629S}) that suggest that the first QSOs are
situated in the most prominent dark matter haloes and are surrounded
by a large number of fainter galaxies.  Although the redshift
discrimination offered by our photometric selection is not very
accurate, the chance of finding a random projection of seven galaxies
at $z\sim6-7$ is quite unlikely given their overall low number
density. We speculate that we are seeing the most distant
proto-cluster centered on a massive QSO at $z=6.43$.

\subsection{Assessing the significance of the overdensity}

\begin{figure}[htbp] 
   \centering
   \includegraphics[width=2in,angle=-90]{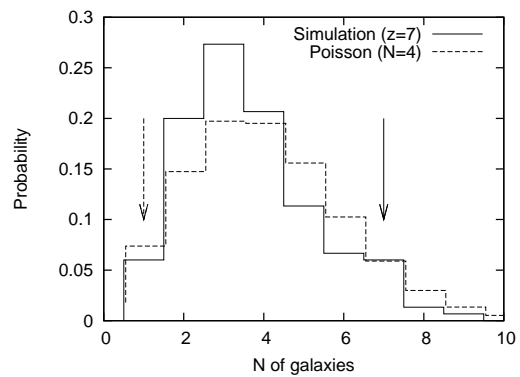} 
   \caption{The probability distribution of how many galaxies would be
     found within the field of view of Suprime-Cam using our selection
     criteria.  The solid histogram is based on the simulation of
     galaxy formation at $z=7$ while the dashed histogram is the
     poisson distribution assuming the average value is $(7+1)/2=4$.
     The solid and the dashed arrows indicate the number of galaxies
     in the QSO field and the SDF comparison field, respectively.}
   \label{fig:prob}
\end{figure}
In this subsection, we assess how significant the overdensity is. If
we assume a Poissonian distribution, the probability of finding seven
objects when one expects one (based on the SDF counts) is less than
0.01\%.  However this is not a fair statistic since we have only one
comparison field (the SDF) which has a similar size as the QSO field.
If we naively calculate the average number of counts expected from the
two fields combined, (7+1)/2, we expect 4 counts on average. The
Poissonian probability to find 7 objects is thus $\sim$11\%. However,
the chance of finding one object in one field and seven in the other is much lower. 
This can be quantified by calculating the expected
number of counts that maximizes the probability of finding seven objects in one
field and one in the other.  Out of 10,000 Monte Carlo realizations we
find 0.40\% of such cases, i.e., the significance of this overdensity
is 99.6\%.

The above estimates are based on the oversimplification that galaxies
are not clustered.  Here we evaluate our results based on the cosmic
variance determined from a cosmological simulation.  The expected
number of galaxies is predicted from the cosmological smoothed
particle hydrodynamics (SPH), cold dark matter (CDM) model of
\citet{Choi2009,Choi2010}. The simulated galaxies have appropriate
colors and magnitudes, allowing us to make a direct comparison to our
observation. In order to simulate galaxies at $z=7$, we have applied
(i) a bright magnitude cut at $z_{R}>24$, and (ii) a detection
completeness.  Then, we count how many galaxies would be found in the
field of view of Suprime-Cam.  To evaluate the completeness of our
detection procedure as a function of magnitude, we add artificial
objects to the $z_{R}$ image with all detected objects masked.  The
artificial objects are modeled by a moffat function with a half light
radius of 0.9 arcsec in the magnitude range of 24 to 25.46.  We used
the IRAF task ``noao.artdata.mkobjects'' for this procedure.  Then, we
repeat the object detection and measure the success rate.

After applying the completeness and magnitude cuts to the simulated
data, we have a realistic sample that can be compared to our
observational data.  By randomly sampling the Suprime-cam size field of view
from the simulation, we measured the galaxy frequency distribution
(Fig. \ref{fig:prob}).  Based on this simulation, we found that the
probability to find one pointing with seven galaxies and another with
one is less than 0.4\%, in good agreement with the results from the
Poissonian statistics detailed above.  Our results suggest that it is
hard to explain the overdensity by a chance coincidence of galaxies
due to cosmic variance.

\begin{table}[tdp]
\begin{center}
\caption{The properties of the $z\sim 6.4$ candidate galaxies detected in the QSO field.}
\begin{tabular}{ccccccc}
\hline
id	&	$\alpha_{2000}$	&	$\delta_{2000}$	&	$z_{R}$		& $i'-z'$	&	$z'-z_{R}$	&	FWHM\\
\hline
\hline
1	&	23:29:20.56	&	-02:54:29.5	&	24.62	&	2.03	&	0.44	&	0.63\\
2	&	23:29:24.10	&	-03:07:34.3	&	24.64	&	$>1.93$	&	0.54	&	1.15\\
3	&	23:29:28.56	&	-02:57:38.4	&	24.69	&	$>1.98$	&	0.52	&	0.55\\
4	&	23:28:48.69	&	-03:06:23.3	&	24.88	&	$>2.08$	&	0.48	&	0.58\\
5	&	23:28:58.68	&	-03:12:11.6	&	24.98	&	$>2.02$	&	0.41	&	0.64\\
6	&	23:30:04.24	&	-02:56:54.6	&	25.00	&	$>1.97$	&	0.38	&	0.69\\
7	&	23:29:31.52	&	-03:11:05.0	&	25.11	&	$>1.93$	&	0.32	&	1.02\\
\hline
\end{tabular}
\label{tab:objectlist}
\end{center}
\end{table}

\subsection{Spatial distribution of the LBGs}

In this subsection, we discuss the peculiar spatial distribution of the LBGs.
\begin{figure}[tbp]
   \centering
   \includegraphics[width=2.8in,angle=-90]{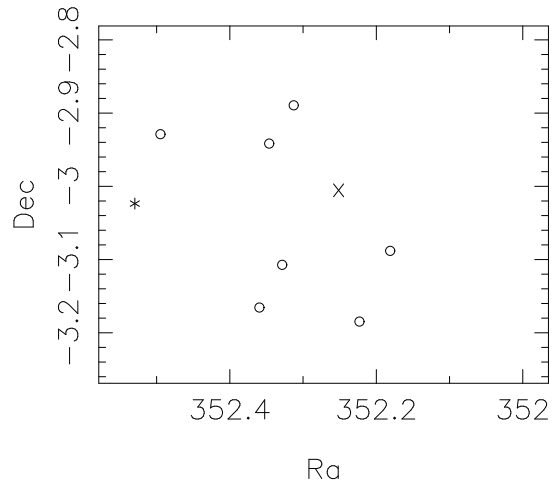} 
   \caption{Spatial distribution of $z>6.4$ LBGs (circles), CFHQ
     J2329-0301 at $z=6.43$ (cross) and a newly identified QSO
     candidate (asterisk).}
   \label{fig:spatial}
\end{figure}
Fig. \ref{fig:spatial} shows the spatial distribution of the $z\sim
6.4$ galaxies.  The cross located at the center indicates the position
of CFHQS J2329-0301, and circles represent the $z\sim 6.4$ galaxies.
It appears that while these galaxies are clustered around the QSO,
none are in its direct vicinity.  (the closest galaxy being already at
a (projected) distance of $\sim$2 Mpc from the QSO).  To quantify the
significance of this effect, we show a histogram of the angular
distances from the QSO to each of the galaxies in
Fig. \ref{fig:histogram} (red solid histogram).  The distribution
indeed suggests that there is a deficit of galaxies at $<$2 Mpc, while
typical clustered distributions are expected to peak at the
center. The blue dotted line in Fig. \ref{fig:histogram} is based on
the positions of randomly distributed objects (but avoiding the masked
regions just as in the real data). Comparing the two histograms
suggests that in the QSO field the number of galaxies at the projected
distance from 2 to 4 (physical) Mpc/h is indeed larger than that
expected from a random distribution.  We perform a Kolmogorov-Smirnov
test by calculating the following value:
\begin{eqnarray}
	D=\max_{i}|R_{i}-M_{i}|,
\end{eqnarray}
where $R_{i}$ is the number of galaxies in the $i$-th bin and $M_{i}$
is the number in the same bin as given by the randomly distributed
galaxies.  We construct a histogram of the $D$ statistic using 10000
times realizations, finding that the distribution in the QSO field
differs from the random distribution at a 97\% confidence level.  If
we just compare the peak at 3 Mpc, it has a 1.7 $\sigma$ excess over the
random distribution.

\begin{figure}[tbp] 
   \centering
   \includegraphics[angle=-90,width=3.2in]{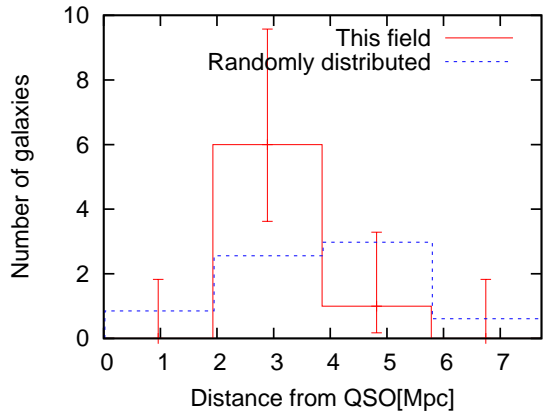} 
   \caption{Histogram of radial distances (in projected, physical Mpc)
     from the QSO to each of the galaxies (red line). The blue dashed
     line was derived by taking the average of 100 realizations of a
     randomly distributed population having the same number of
     galaxies. Error bars indicate the 1 sigma poisson error using
     Eqn. (9) and Eqn (13) from \citet{1986ApJ...303..336G}. We find a
     $1.7 \sigma$ excess in the number of galaxies at a distance of
     $\sim3{\rm Mpc}/h_{70}$. Interestingly, this distance is similar
     to the size of the HII region around $z\sim6$ QSO
     \citep{2005ApJ...628..575W}.}
   \label{fig:histogram}
\end{figure}

\subsection{Evidence for additional QSOs?}  

Finally, we search for additional QSOs in the field by applying a set
of color-color criteria tuned to the colors expected for QSOs at
$z\sim6.4$.  Since a QSO continuum decreases with wavelength blue-ward
of Ly$\alpha$, a QSO will have $i'-z'>1.9$ and $z'-z_{R}<0.3$
(Fig. \ref{fig:spectrum}).  Note that this is a different (bluer) cut
than that used for galaxies.  We apply these criteria to all objects
detected at $>$3$\sigma$ in all bands and having a FWHM smaller than
0.65 arcsec (equal to ${\rm FWHM}+\sigma_{\rm FWHM}$).  We found two
objects (see Table \ref{tab:qsolist}).  One of these is CFHQS
J2329-0301, showing that our QSO selection works.  Although the
redshift and nature of the second object must be confirmed with
spectroscopy, it may be another QSO associated with the structure of
galaxies surrounding CFHQS J2329-0301. If so, this would add
supporting evidence that this field is relatively overdense.  The
position of the QSO candidate is indicated by the asterisk in
Fig.\ref{fig:spatial}.

\begin{table}[tdp]
\begin{center}
{\tiny 

\caption{A new QSO candidate found based on color-color selection.}\label{tab:qsolist}
\begin{tabular}{ccccccc}
\hline
id	&	$\alpha_{2000}$	&	$\delta_{2000}$	&	$z_{R}$		& $i'-z'$	&	$z'-z_{R}$	&	FWHM\\
\hline
\hline
8	&	23:30:14.95	&	-03:03:17.1	&	19.45	&	3.26	&	0.14	&	0.60\\
\hline
CFHQS	&	23:29:08.28	&	-03:01:58.2	&	21.76	&	4.11	&	-0.39	&	0.57\\
\hline
\end{tabular}
}

\end{center}
\end{table}%

\section{Discussion}

We found an overdensity of LBGs around the QSO with a 99.6\%
significance. We furthermore found evidence for a ring-shaped
distribution, albeit at the $<$2 sigma level.  Although the physical
interpretation must await verification using deeper data or
spectroscopic follow-up, below we will discuss possible physical
interpretations under the assumption that our results are significant
and represent real structure surrounding the QSO.

\subsection{Comparison with previous work}

At lower redshift, the environments of QSOs and fainter AGN have been
extensively studied.  At $0.05 <z<0.095$, \citet{2003ApJ...597..142M}
found that the fraction of galaxies with an AGN is independent of the
local galaxy density, in a stark contrast to both star-forming and
passive galaxies that show an environmental dependence
\citep[e.g.,][]{2003PASJ...55..757G}.  \citet{2009A&A...501..145L}
found an underdensity of bright galaxies at a few Mpc scale from
nearby QSOs at $0.078<z<0.172$.  \citet{2004MNRAS.353..713K} found
that at fixed stellar mass the number of galaxies that host active
galactic nuclei (AGNs) with strong [O III] emission indicating strong
AGN activity is twice as high in low-density regions compared to high
density regions at $0.04 < z < 0.06$.

However the situation is different at higher redshifts.  A Keck survey
of fields centered on known $z>4$ QSOs found excesses in the number of
companion galaxies \citep{1999ASPC..193..397D}.
\citet{2007ApJ...663..765K} found that LBGs without Ly$\alpha$
emission form a filamentary structure near a QSO at $z\sim5$, while
Ly$\alpha$ emitters are distributed around it but avoid it within a
distance of $\sim$ 4.5 Mpc.  \citet{2004Natur.427...47M} also found
that LBGs are concentrated around a luminous radio galaxy at
$z\approx4.1$ previously found to be associated with an overdensity of
$\sim$30 spectroscopically confirmed Ly$\alpha$ emitters
\citep{2007A&A...461..823V,2002ApJ...569L..11V}.  In
\citet{2006ApJ...637...58O,2008ApJ...673..143O} it was shown that the
environments of some radio galaxies at $z=4-5$ appear richer than the
average field at $\sim3-5\sigma$ significance based on a detailed
comparison with GOODS. However, these studies found no difference
between the physical properties of galaxies in protoclusters compared
to those in the field.

The difference between the low-$z$ and high-$z$ environments of AGN
perhaps stems from the different halo masses that host them. Due to
the flux limited nature of observational surveys, high-$z$ QSOs are
much more luminous than local AGNs, and thus, presumably embedded in a
more massive halo.  Indeed, a generic expectation in most models of
galaxy formation is that the most massive density peaks in the early
universe (such as QSOs and massive galaxies) are likely to be strongly
clustered \citep{1984ApJ...282..374K,1988MNRAS.230P...5E}.  In the
hierarchical formation and evolution scenario of galaxies and QSOs
\citep{2000MNRAS.318L..35H}, luminous QSOs are located in rare
overdense regions.  This is why high-$z$ QSOs have been used as
beacons to search for high density regions
\citep{2006MNRAS.371..786C}.  \citet{2005MNRAS.356..415C} and
\citet{2007AJ....133.2222S} found an increasing clustering of QSOs
with redshift.

Surprisingly, at $z>6$ the situation appears to change again; luminous
QSOs are not necessarily found in strong density peaks.
\citet{2009ApJ...695..809K} studied the number densities of
$i$-dropout objects around 5 SDSS $z\sim$6 QSOs using the
HST/ACS. They found an overdensity in two fields and underdensity in
two fields.  \citet{2006ApJ...640..574Z} and
\cite{2005ApJ...622L...1S} found an overdensities around the QSOs
SDSSJ0836+0054 ($z=5.82$) and SDSS J1030+0524 ($z=6.28$).  However,
these overdensities do not appear to be as magnificent structures as were found at lower
redshift, or even in some random regions at $z\sim5-6$ \citep[e.g.,
][]{2004ApJ...611..685O,2007ApJ...663..765K,2008ApJ...677...12O}.

One of the major problems of previous work at $z\sim$6 is the
relatively small FoV of the HST/ACS.  The 200''$\times$200'' field can
only probe a region of 1 Mpc $\times$ 1 Mpc at $z\sim$6, and thus may
easily miss any larger structures such as found in this work. Computer
simulations also predict that the largest structures present at
$z\sim6$ span several tens of Mpcs\citep{2009MNRAS.394..577O}, while
LFs of LBGs show a large field-to-field variation
\citep{2004AJ....127..563H,2009ApJ...706.1136O}.  These results
suggest that previous non-detections of overdensities around QSOs may
need to be re-examined using a larger field of view or deeper
observations. The difference in the FoV is perhaps the primary reason
why no previous work has detected a highly significant overdensity
around any of the $z\sim$6 QSOs.  Our positive detection may have been
facilitated by the large field of view that allowed us to investigate
the structure at scale of $\sim$3 Mpc.

 
 \begin{figure}[htbp]
    \centering
    \includegraphics[width=3in]{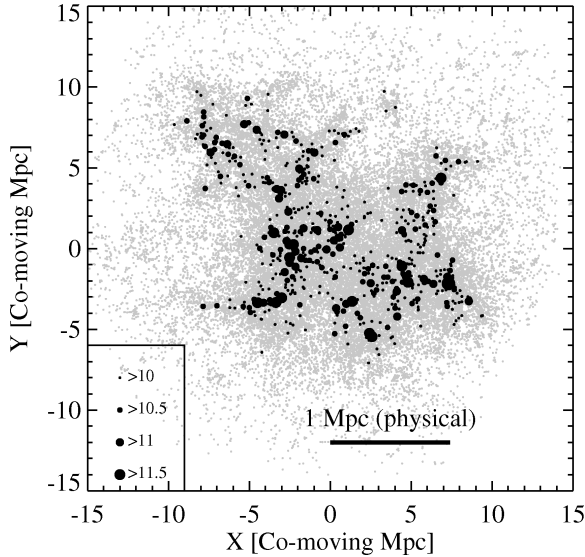} 
    \caption{Illustration of a massive protocluster region selected
      from a large cosmological $N$-body simulation. The protocluster
      region shown is the progenitor of the most massive cluster found
      in the {\it Millennium-II Simulations} (corresponding to a
      $M\simeq10^{15}$ $M_\odot$ cluster at $z=0$). Symbols show the
      (projected) distribution of dark matter halos having masses as
      indicated in the figure legend (values indicate the logarithm of
      the halo mass in $M_\odot$). The scale bar indicates a physical radius of 1
      Mpc.}
    \label{fig:simdistribution}
 \end{figure}
 
 For an illustration of what such a region at $z\gtrsim6$ might look
 like, we show an example of a large protocluster selected from the
 {\it Millennium-II} cosmological $N$-body simulations
 \citep{2009MNRAS.398.1150B}.  Figure \ref{fig:simdistribution} shows
 the (projected) spatial distribution of strongly clustered dark
 matter halos associated with a protocluster at $z=6.2$.  This
 protocluster is the progenitor of the most massive cluster found in
 the {\it Millennium-II} simulations, and corresponds to a
 $M\simeq10^{15}$ $M_\odot$ cluster when evolved to $z=0$. Figure
 \ref{fig:simdistribution} shows, at least qualitatively, that we may
 expect significant enhancements in the density of the galaxies that
 are hosted by the dark matter halos shown, provided that QSOs indeed
 trace protoclusters. A more quantitative analysis of the surface
 density of $z$-dropout galaxies near QSOs expected in the simulations
 is needed for a proper comparison.

\subsection{Size of the HII region}

Next, we compare our finding of an overdensity of LBGs around a QSO at
a characteristic scale of 3 Mpc to spectroscopic measurements of HII
regions around QSOs.

It has been demonstrated through spectroscopic observations that there
exist large, ionized regions around luminous QSOs.  These are
sometimes called HII regions, Stromgren spheres, or highly ionized
near zones.  In this work, we will refer to them as HII regions.  In
order to define the size of the HII region,
\citet{2006AJ....132..117F} proposed a definition as a point in the
spectra where the Ly$\alpha$ transmission first drops to $T<0.1$ for
spectra binned in 20\AA~ pixels.  The CFHQS J2329-0301 transmission
drops at $T<0.1$ first at 3.6 Mpc, then again at 6.3 Mpc
\citep{2007AJ....134.2435W}.  Interestingly, the size of this
spectroscopically measured HII region is comparable to the possible
ring shape distribution of LBGs around CFHQS2329-0301
(Fig. \ref{fig:spatial}).  Note that the size of the spectroscopic HII
region is expected to be larger ($\sim$10 Mpc) for the SDSS QSOs due
to their higher luminosities (they are brighter by by $\sim2$ mag).
In addition to the small FoV of HST/ACS, this may be an additional
reason, why no significant overdensity of LBGs has been found around
SDSS QSOs.  To observe an overdensity of galaxies at a scale of $>$10
Mpc (such as expected around the luminous SDSS QSOs at $z\sim6$), one
needs multiple FoVs even with the Suprime-Cam.  We conclude that the
size of the HII region is consistent with the apparent lack of LBG
candidates closest to the QSO.

\subsection{Possible Physical Mechanisms}

If our detection of the lack of LBGs near the QSO is real, then, what
created the observed paucity of galaxies within 3 Mpc from the QSO?
One possibility is that the intense emission of ionizing radiation
associated with QSOs ionizes the surrounding IGM and may even
photo-evaporate the gas in neighboring dark halos before it has the
opportunity to cool and form stars. In this scenario, QSOs would
suppress galaxy formation in their vicinity. One would then observe a
paucity of galaxies near a QSO despite the underlying excess of dark
matter halos.  \citet{2004MNRAS.348..753S} presented the first
theoretical simulations of the gas dynamics coupled with radiative
transfer, showing that an ionizing source that emits 10$^{56}$ photons
s$^{-1}$ (appropriate for a QSO) can indeed photoevaporize minihalos
of $\sim$10$^7 M_{\odot}$ (that would have been able to form small
galaxies) on 100--150 Myr timescales \citep[but
see][]{1999ApJ...523...54B,2005ApJ...628..575W}.  Compared with our
observational results of finding a ``ring''-shaped distribution of
LBGs, the central QSO may have suppressed the formation of surrounding
galaxies thereby creating a paucity of galaxies in its vicinity but
still having an overdense region beyond the inner, ionized region.
This may explain the observed results, at least qualitatively.

However, \citet{2007ApJ...663..765K} quantitatively estimated that
such UV radiation from QSO can suppress galaxy formation or not.
According to their simulation, QSO UV radiation can suppress star formation
in a halo with $M_{vir}<10^9M_{\odot}$, while a halo with
$M_{vir}>10^{11}M_{\odot}$ is unaffected. 
Although mass estimate of our $z\sim6.4$ LBG are uncertain
because we do not have deep near-infrared data, considering bright magnitude,
they are likely to be massive galaxies with $M_{vir}\sim10^{11-12} M_{\odot}$.
 If so, QSO UV radiation is not strong enough to create the central deficit we observed.


Can we find other theoretical predictions that could explain the
observations?. The simulation of \citet{2007ApJ...663..765K} assumed
star formation in spherical dark matter halos.  They found that the
impact of photoionization is greater if that star formation is taking
place in a disk or in substructures (which is perhaps likely to be the
case). If the QSO formed at a much earlier time than the surrounding
galaxies, this scenario might be able to suppress the formation of
galaxy seeds at the time when their masses were still sufficiently
small.

Another plausible scenario is that the IGM surrounding the QSO may have been ionized by galaxies long before the QSO turned on.  
It has been suggested that because luminous QSOs are expected to form
in rare overdense regions, the surrounding IGM had already been
pre-ionized by galaxies \citep{2005ApJ...620...31Y}.  According to the
simulation of \citet{2007MNRAS.381L..35B}, the neutral hydrogen
fraction, $f_{HI}$, near a QSO is estimated to be
$f_{HI}<0.3$. Therefore, QSO radiation is emitted into a substantially
pre-ionized IGM. In this case, galaxies that formed before the QSO
started emitting its strong UV radiation may still be present in the
vicinity of the QSO. Because such galaxies are likely to have ceased their star formation, they would not be detected by our LBG technique.  

In an overdense region, the QSO host galaxy is likely to be the most
massive galaxy that formed the earliest.  According to
\citet{2005ApJ...620...31Y}, QSO host galaxies experiencing rapid star
formation at a rate of $\sim$3000$M_{\odot}$ yr$^{-1}$ combined with
the radiation field emitted by the QSO itself can produce enough
photons to ionize a large HII region. In order to check whether we see
any evidence for the presence of such a massive host galaxy, we have
performed a careful PSF subtraction on our images. The residuals
indicate the detection of a large host galaxy by SED fitting
($R_e$=11kpc, $\lesssim10^{10}M_{\odot}$) associated
with the QSO CFHQS J2329-0301 \citep{2009MNRAS.400..843G} with
evidence of extensive star formation based on its extended rest-frame
UV flux. The presence of a massive host galaxy may thus support a scenario in which 
the QSO and its host galaxy evolve together, suppressing 
galaxy formation in their vicinity.

Alternatively, it has also been suggested by
\citet{2008MNRAS.391.1961D} that in order to facilitate the formation
of a supermassive black hole by $z\sim6$ in the first place, it may be
required to have a rare pair of dark matter halos in which the intense
UV radiation from one halo prevents fragmentation of the other, so
that the gas collapses directly into a supermassive black hole, and
explaining the lack of galaxies forming in its vicinity as observed in
our field.
Consistent with this scenario, it is interesting to note that we found
a second QSO candidate located near the eastern edge of the field (see
$\S$4.4). We speculate that this could be a QSO associated with the
adjacent (second) halo that ionized the original halo.

Last, it has been suggested that a $z\sim6$ QSO is likely to have
experienced multiple mergers in order for its black hole to grow to a
mass of $\sim 10^9 M_{\odot}$. An example of this scenario is given by
the simulation of \citet{2007ApJ...665..187L}, which predicts that the
host galaxy of a $z\sim6$ QSO may have experienced seven mergers with
mass ratios of 4:1 or greater (see their Fig. 3). The QSO in our field may thus have
merged with all the galaxies in its direct vicinity, explaining the
peculiar spatial distribution that we find. This is also consistent
with the discovery of the large host galaxy associated with the QSO
\citep{2009MNRAS.400..843G}.


In summary, although our detection of substructure in the LBG
distribution near the QSO is rather weak ($\lesssim2\sigma$), studies
of the spatial distribution of LBGs around QSOs at $z\sim6$ are very
important for testing numerous of the interesting physical scenarios
related to the co-evolution of QSOs and galaxies as discussed above.
Therefore, it is important that our conclusions are verified using
deeper, multi-wavelength imaging and spectroscopic observations that
may detect additional galaxies missed by our current selection, such
as galaxies that are below our (UV) detection limits, star-forming
galaxies that are heavily obscured, or galaxies in which star
formation has ceased.  Also, it will be important to extend the
analysis performed here to other fields to obtain good statistics on
the typical structures associated with QSOs at this extreme redshift.

\section{Conclusions}

Taking advantage of the large field of view (34'$\times$27') and the
new red-sensitive fully depleted CCDs recently installed on the
Subaru/Suprime-Cam, we have investigated the large scale
environment around the most distant QSO studied to date.

Our findings are as follows. The number of candidate LBGs at
$z\sim6.4$ is 7 times larger than that in a comparison field (the
Subaru Deep Field), suggesting that the QSO field hosts an overdense
region. We estimate that the probability that this overdensity is a
chance coincidence is less than 0.4\% based on either using simple
Poissonian statistics, or on a comparison with a cosmological SPH
simulation of $z\sim7$ galaxies.

We find evidence for a non-uniform distribution of the LBGs in a
``ring-like'' shape surrounding the QSO at a radius of $\sim$3
physical Mpc, i.e., galaxies are overdense around the QSO, but at the
same time they avoid the very center near the QSO (see
Figs. \ref{fig:spatial} and \ref{fig:histogram}).  A KS-test shows
that the radial distribution of LBGs in Fig. \ref{fig:histogram} is
different from random at a 98\% significance level.  The distance of 3
Mpc is comparable to the size of the HII region around QSOs at
$z\sim6$ \citep{2005ApJ...628..575W}. Possible physical explanations
of such a central deficit of galaxies include the suppression of
galaxy formation due to the strong UV radiation field of the QSO.
However, because the significance of our detection of a non-uniform
distribution of LBGs around the QSO is low, it is important to verify
these results with deeper imaging data and/or spectroscopy.

Our findings show that QSOs at $z\sim6$ may indeed be embedded in the
densest regions of the early universe, provided that they are observed
on significantly larger scales compared to previous studies that used
the relatively small field of view provided by HST.

\section{Acknowledgments}

We thank the anonymous referee for careful reading of the manuscript
and many insightful suggestions.  We are greatful to K. Shimasaku and
M. Ouchi for providing $z_{R}$ filter, valuable comments and a careful
reading of this paper and also thank N. Asami for providing valuable
comments.  We thank Jason Jaacks, Jun-Hwan Choi, and Kentaro Nagamine
for providing us with the results of their cosmological hydrodynamic
simulation.

Y.U. acknowledges grant aid for large storage system from the
Department of Astronomical Sciences of the Graduate University for
Advanced Studies (SOKENDAI).  T.G. acknowledges financial support from
the Japan Society for the Promotion of Science (JSPS) through JSPS
Research Fellowships for Young Scientists.

The Millennium-II Simulation databases used in this paper and the web
application providing online access to them were constructed as part of
the activities of the German Astrophysical Virtual Observatory.

The authors wish to recognize and acknowledge the very significant
cultural role and reverence that the summit of Mauna Kea has always
had within the indigenous Hawaiian community.  We are most fortunate
to have the opportunity to conduct observations from this sacred
mountain.

\end{document}